\documentclass[fleqn,10pt]{wlscirep}
\title{Additive manufacturing of magnetic shielding and ultra-high vacuum flange for cold atom sensors}

\author[1]{Jamie Vovrosh}
\author[1,3]{Georgios Voulazeris}
\author[1]{Plamen Petrov}
\author[2]{Ji Zou}
\author[2]{Youssef Gaber}
\author[3]{Laura Benn}
\author[3]{David Woolger}
\author[2]{Moataz M. Attallah}
\author[1]{Vincent Boyer}
\author[1]{Kai Bongs}
\author[1,*]{Michael Holynski}
\affil[1]{School of Physics and Astronomy, University of Birmingham, Birmingham, B15 2TT, UK.}
\affil[2]{School of Metallurgy and Materials, University of Birmingham, Birmingham, B15 2TT, UK}
\affil[3]{Magnetic Shields Limited, Staplehurst, UK}

\affil[*]{m.holynski@bham.ac.uk }

\begin{abstract}
Recent advances in the understanding and control of quantum technologies, such as those based on cold atoms, have resulted in devices with extraordinary metrological performance. To realise this potential outside of a lab environment the size, weight and power consumption need to be reduced. Here we demonstrate the use of laser powder bed fusion, an additive manufacturing technique, as a production technique relevant to the manufacture of quantum sensors. As a demonstration we have constructed two key components using additive manufacturing, namely magnetic shielding and vacuum chambers. The initial prototypes for magnetic shields show shielding factors within a factor of 3 of conventional approaches. The vacuum demonstrator device shows that 3D-printed titanium structures are suitable for use as vacuum chambers, with the test system reaching base pressures of $5 \pm 0.5 \times 10^{-10}$ mbar. These demonstrations show considerable promise for the use of additive manufacturing for cold atom based quantum technologies, in future enabling improved integrated structures, allowing for the reduction in size, weight and assembly complexity.
\end{abstract}
\begin{document}

\flushbottom
\maketitle
%
%
\thispagestyle{empty}

\section*{Introduction}

Quantum technologies utilising atom clouds are highly promising tools for creating ever more sensitive devices with applications in a vast array of areas ranging from geophysical type applications \cite{BODDICE2017149,dowling2003quantum} to satellite independent navigation \cite{keil2016fifteen}. The exceptional performance of lab based systems \cite{Kovachy2015, BloomClock, Sander:12} has lead to recent work focused on transforming lab based atomic systems into compact transportable versions \cite{hinton2017portable,wu2014investigation,hauth2013first,menoret2011dual}. To produce ever more compact and transportable devices the latest advances in micro-manufacturing technology are being used including waveguide writing \cite{politi2008silica} and reactive ion beam etching \cite{sewell2010atom}. Using these manufacturing methods has allowed for the miniaturisation of quantum technologies such as integrated atom chip based systems \cite{PhysRevLett.110.093602,PhysRevA.96.033636}.

An emerging technology capable of allowing further miniaturisation of cold atom based sensors is additive manufacturing, such as 3D-printing. The freedom of design \cite{Wonderfulwidgets} offered during production by 3D-printing allows for rapid development of complex, individually bespoke components and the potential to tune material properties during production\cite{TheEconomist}.
3D-printing has been used for a variety of applications ranging from fast compact laser shutters \cite{doi:10.1063/1.4937614} to the realisation of complex magnetic field configurations \cite{Ling,doi:10.1063/1.4964856}. The focus of 3D-printing technologies in regards to cold atom based sensors, thus far has been on magnetic field generation \cite{zhou2017design, saint20173d}, while little attention has been given to environmental isolation. Here we report on two demonstrators of crucial environmental isolation technologies needed for the realisation of portable and compact quantum technologies developed with 3D-printing, namely magnetic shielding and vacuum components.

Magnetic shielding is an essential component of atom based quantum technologies necessary to provide a suitable magnetic environment and enable sensitive measurements. Currently the best available materials for magnetic shielding are soft magnetic alloys, such as mu-metal \cite{tong2016advanced}. Despite mu-metal shields being widely used, they are characterised by a relatively high weight and the inflexibility to adapt to more complex geometries, due to manufacturing limitations. While simple shapes such as cylinders can be easily manufactured, linking shields together without loss of performance is challenging \cite{doi:10.1063/1.4720943}. Existing shielding is thus heavy and bulky, limiting the advancement of quantum technology towards portable and miniaturised systems. The majority of production is currently realised through hand machining in workshops which limits the ability to create complicated geometries. In contrast, by using 3D-printing  it is possible to design bespoke complex shielding for each application. However to achieve the best possible magnetic shields the additive process used to create the shields need to be optimised to reduce cracking and porosity in the printed structure, which reduces the effectiveness of the magnetic shields. Combining this with post processing of the 3D-printed structure, the ability to produce magnetic shields with complex compact  and light weight geometries closer to the structure to be shielded should be realised.

Vacuum technology is another prominent component of cold atom systems, used to isolate atoms from background gasses.  For example, a typical measurement rate in an atom interferometer is on the order of 1 Hz. In order to not be limited by background gas collisions, this requires a pressure of the order of $10^{-9}$ mbar \cite{PhysRevLett.90.160401}. For compact quantum sensors it is ideal that ultra high vacuum environments are achieved and maintained over long time periods. Since 3D-printing is an additive process, it is possible that small voids and leak channels can form in the material during the printing process. These voids could trap gases which would later vent slowly into the vacuum, making the part unsuitable for ultrahigh vacuum environments \cite{das1998producing}. Selection of the material which components are made out of has been shown to reduce these issues in 3D-printed components inside vacuums \cite{povilus2014vacuum}. Here we show that by optimising the 3D-printing parameters and materials, 3D-printing techniques can be used to produce vacuum components. The production of 3D-printed vacuum parts allows for the time and cost to build complicated, specialised geometries to be drastically reduced.

Here, for the first time, we report on the design, manufacture and characterisation of both 3D-printed magnetic shielding and vacuum flange demonstrator devices. We evaluate the performance of these demonstrator devices and an equivalent traditionally produced component  of the 3D-printed equivalent.

\section*{\label{sec:Results} Results}

\subsection*{\label{sec:level1} Magnetic Shielding}

\begin{figure}[h]
\centering

\includegraphics[scale=0.4]{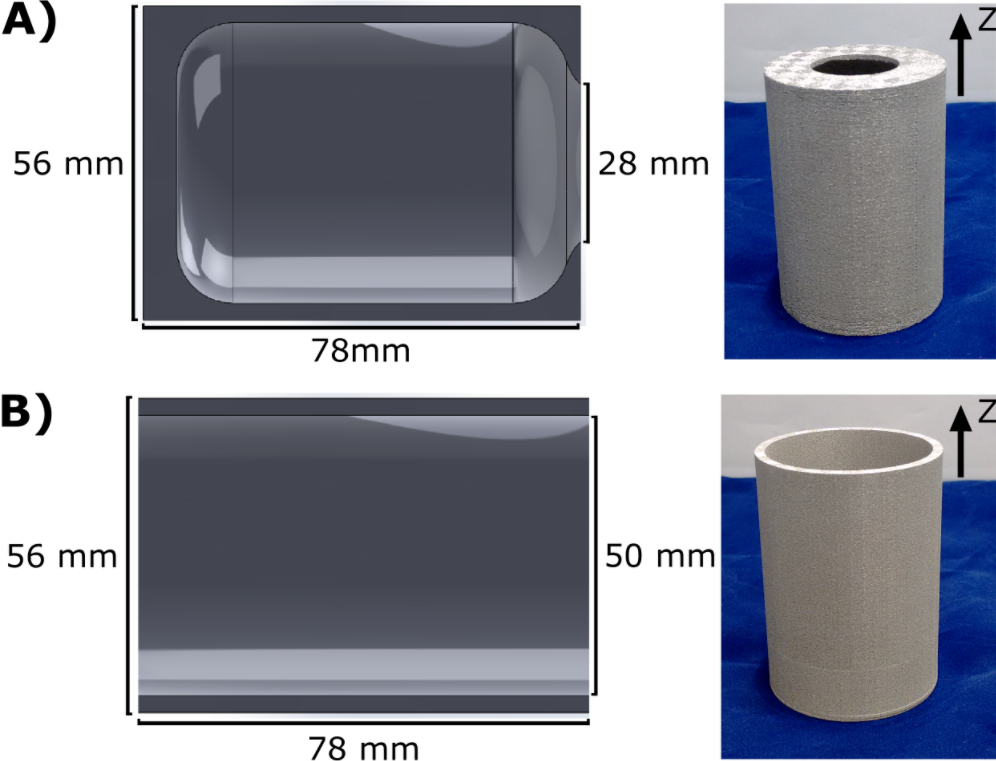}
\caption{The 3D-printed magnetic shield structures used as test pieces. A) A cross section of the Z-Y plane showing the closed ended magnetic shield dimensions and a photo of the 3D-printed structure. B) A cross section of the Z-Y plane showing the open ended magnetic shield dimensions and a photo of the 3D-printed structure.}
\label{fig:MegneticShield}
\end{figure}

To investigate the potential for 3D-printing to be used in the fabrication of magnetic shields the test pieces shown in figure \ref{fig:MegneticShield} were produced. The magnetic shield was designed using computer aided design (CAD) and 3D-printed using the selective laser melting (SLM) process \cite{parimi2014additive,shishkovsky2015comparison}. This method utilises a  high-power scanning laser to locally melt powdered metal, in our case a Ni-5Mo-15Fe alloy (or permalloy-80), allowing for a 3D structure to be built layer by layer. The SLM process allows for optimal customisation and experimentation on the alloy composition, with the printing parameters affecting the micro-structural characteristics of the material, and hence its magnetic properties \cite{mikler2017laser,shishkovsky2016peculiarities}. 

\begin{figure}[h]
\centering
\includegraphics[scale=0.4]{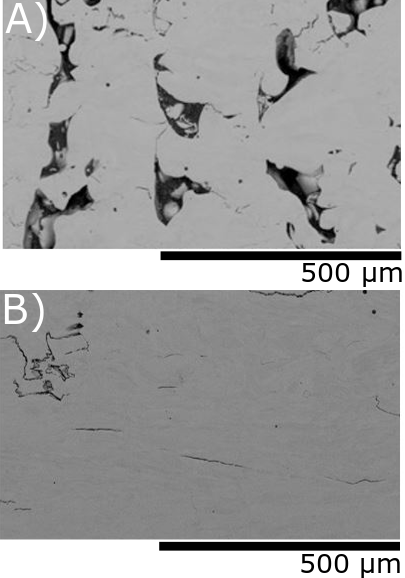}
\caption{Pictures captured by a SEM on different 3D-printed samples revealing structural defects. A) Sample printed under non-optimal parameters with ED $=1.4$ J/mm$^2$ and shows a large number of defects. B) Sample produced with under optimum parameters with ED $= 4.9$ J/mm$^2$ showing minimal defects.}
\label{fig:surfaceimages}
\end{figure}

The defects in terms of voids and cracking can retard and hinder the magnetic domain wall motion \cite{shishkovsky2016peculiarities}. As a result more energy is required to magnetise or demagnetise the material, degrading its permeability. The porosity and cracking within the fabricated structure is a function of the energy density (ED) input to the material surface during the printing process \cite{carter2014influence, paperno2000new}. The energy density is a semi-empirical quantity describing the consolidation heat input given to the powder.  Each material has a threshold level to achieve full consolidation \cite{Carter}.  With the increase in densification and reduction of structural defects, the material is capable of delivering better magnetic shielding. Several samples were prepared with different ED values to minimise such unwanted defects. Each of the samples where characterised by cutting the sample into thin slices which were polished and then examined with a scanning electron microscope (SEM). Pictures where taken from different sections and analysed by imaging software which estimates the percentage of porosity. This method allowed the analysis of the structural uniformity and to see what effect the printing conditions had on the samples produced. Figure \ref{fig:surfaceimages} shows an example of pictures captured by the SEM. The sample shown in figure \ref{fig:surfaceimages} A was printed with ED $=1.4$ J/mm$^2$ and has a structure characterised by a large number of pores with unconsolidated particles and cracks. In contrast, the sample shown in figure \ref{fig:surfaceimages} B was printed under ED $= 4.9$ J/mm$^2$ and showed the best results with negligible porosity and only few cracks.

\begin{figure}[h]
\centering
\includegraphics[scale=0.55]{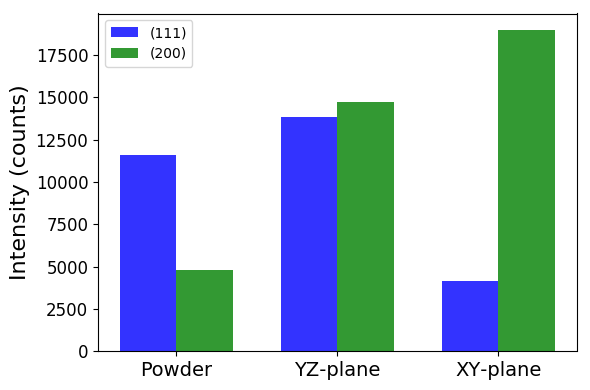}
\caption{Comparison of the results from XRD tests on the unprocessed permalloy-80 powder and the sample produced with ED $= 4.9$ J/mm$^2$ along the XY and YZ-plane, respectively. The z axis is defined as the vertical axis of the sample. Each peak represents the intensity of the detected regions with the same crystallographic orientation. Although both phases are almost equally distributed on the YZ-plane, the (200) orientation dominates in the XY plane.}
\label{fig:XRDMeasurements}
\end{figure}

In addition, phase assemblage and crystallographic texture of the as-fabricated permalloy were characterised using X-ray diffraction (XRD). This revealed a variation in the  crystallographic texture in the two different planes (see figure \ref{fig:XRDMeasurements}). From the XRD pattern, it can be seen that a (001) grain orientation exists in both axes of the magnetic shields. The (001) is the hard axis for an Ni enriched alloy, such as the permalloy-80 used to create the magnetic shields. The hard axis is energetically unfavourable for magnetisation. Therefore presence of the (001) grain orientation will have a negative impact on the shielding factor of the shields \cite{o1999modern}. The nature of powder bed melting process means that grains tend to solidify and grow along the [001] direction, which is the favourable growth direction in Ni \cite{lampman1997weld}. Therefore the powder bed melting process is likely to generate an orientation that is magnetically unfavourable. With further optimisation of the SLM process used there is scope to reduce the percentage of the material with the (001) grain orientation. However for the purposes of creating a magnetic shield the crystallographic properties look favourable in the transverse direction (XY-plane) which is the plane of interest for these experiments.

\begin{figure}[h]
\centering
\includegraphics[scale=0.4]{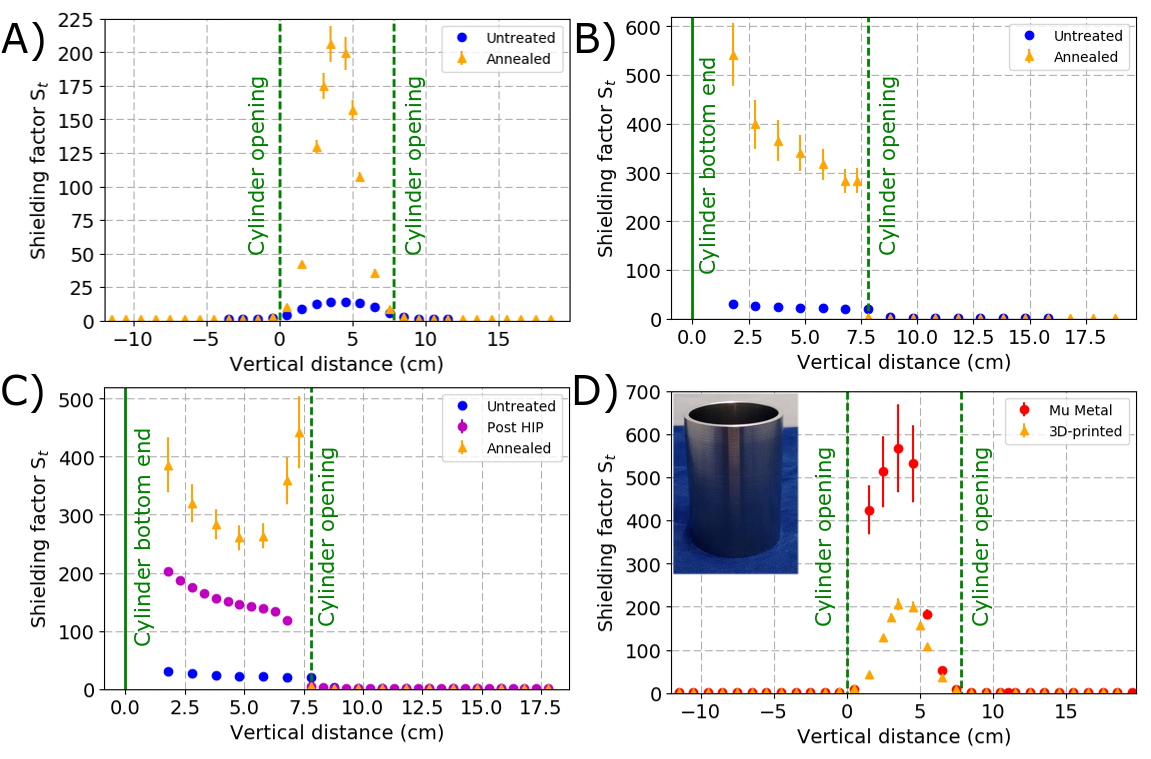}
\caption{Measured shielding factors along the test and control shields longitudinal axes (z axes), under an external magnetic field of 50 $\mu$T in the transverse (Y plane) of the shields. The dashed lines show the position of the open ends of the shields, while the solid lines shows the closed ends. A) Open ended test shield B) Closed ended test shield C) Closed ended test shield which underwent HIP treatment. The increase in shielding factor towards the open end of the test shield has been seen in 3D-printed shields that have not received the HIP treatment and therefore is not an effect of the HIP treatment. It is believed that the rise is an caused by edge effects related to the increased amount of material in that region. D) Comparison of the open ended test shield and the control shield. The inserted image shows the replica mu-metal shield. The mu-metal replica was annealed in pure dry hydrogen under the same conditions as the 3D-printed samples. The maximum magnetic shielding factor measured in the control shield was $568 \pm 102$.
}
\label{fig:OpenField}
\end{figure}

To characterise the magnetic shielding performance the amplitude ratio of the magnetic field without the presence of the shield $B_{out}$, over the residual field measured at the same point after the shield installation $B_{in}$ is used. This ratio is known as the shielding factor $S_{t} = B_{out}/B_{in}$, where the subscript t indicates the magnetising field orientation along the transverse shield axis. The magnetic shielding factors from both before and after annealing the test pieces, when using a test field of 50 $\mu$T magnetic field, can be seen in figures \ref{fig:OpenField} A and \ref{fig:OpenField} B. It can be seen that the maximum magnetic shielding factors obtained before annealing for the open and closed pieces were $14 \pm 0.1$ and $30 \pm 0.3$ respectively. These maximum values were then increased significantly with the application of the annealing process for both the open and closed pieces to $206 \pm 13$ and $542 \pm 64$ respectively. 

To see whether further treatment could improve the performance, heat treatment under hot isostatic pressing (HIP) was also investigated. HIP is often performed during manufacture to release internal stresses within the materials used \cite{g}. The effect of applying HIP treatment prior to annealing on a 3D-printed test piece can be seen in figure \ref{fig:OpenField} C. The details of the HIP cycle parameters can be found in the Methods section. It can be seen that the shielding factor increased by a factor of 7 after the HIP treatment, reaching a value of $S_{t} \approx 150$ at the cylinder central region. This factor was then doubled after annealing, reaching a shielding factor of $S_{t} \approx 260$, approximately at the centre of the shield. It can be seen in figure \ref{fig:OpenField} B the test piece which did not undergo HIP treatment demonstrated very similar performance, suggesting that although HIP as an intermediate step is important for optimising structural integrity it might not be necessary for magnetic shielding applications. In that case, opting to follow only one heat treatment instead of two, could reduce potential production cost in the future. Alternatively, it is possible that further optimisation of the HIP parameters may also lead to further improvements.

The performance of the 3D-printed open ended test piece was compared to that of a mu-metal control shield with the same dimensions, fabricated using standard techniques. These were both processed using the same procedure, which is optimised based upon the mu-metal shield. A comparison of the magnetic shielding factors can be seen in figure \ref{fig:OpenField} D. The shielding factor of the test piece is found to be a factor of 2.75 smaller than the control shield, showing that 3D-printing is a viable method in which the creation of high performance magnetic shields is possible. In particular, this would be of  benefit in the creation of complex geometries, which may allow for substantial geometric improvements in the shield design to enable considerably better shielding per unit weight. Furthermore, optimisation of the fabrication and post-processing may yield further improvements.

\subsection*{\label{sec:level1} Vacuum components}

Many materials that are commonly used in ultra high vacuum (UHV) are also 3D-printable materials, such as silver, gold, stainless steel and titanium. The low vapour pressures of metals make them ideal for vacuum applications, however this is not the only consideration in determining UHV compatibility. The processes used in 3D-printing metals could potentially give rise to several adverse effects, such as introducing trapped gases resulting in virtual leaks, and micro-leak channels. In addition, the resulting surface roughness and porosity of the printed part can greatly increase effective surface area and therefore out gassing load. These problems must be ruled out if 3D-printed materials are to be used to create UHV systems. Additionally it is important to determine if any special surface cleaning or passivation steps are necessary before 3D-printed vacuum parts can be used. Currently the vacuum compatibility of 3D-printed materials has been shown for small components that sit inside a vacuum system, such as for Al-Si10-Mg, titanium and silver \cite{saint20173d, povilus2014vacuum, gans2014vacuum} and to make low vacuum KF vacuum parts \cite{1742-6596-874-1-012097}. However here we will show that 3D-printed parts can be used to construct high vacuum chambers, through acting as the vacuum wall and sealing to peripheral components, using indium sealing.

\begin{figure}[h]
\centering
\includegraphics[scale=0.4]{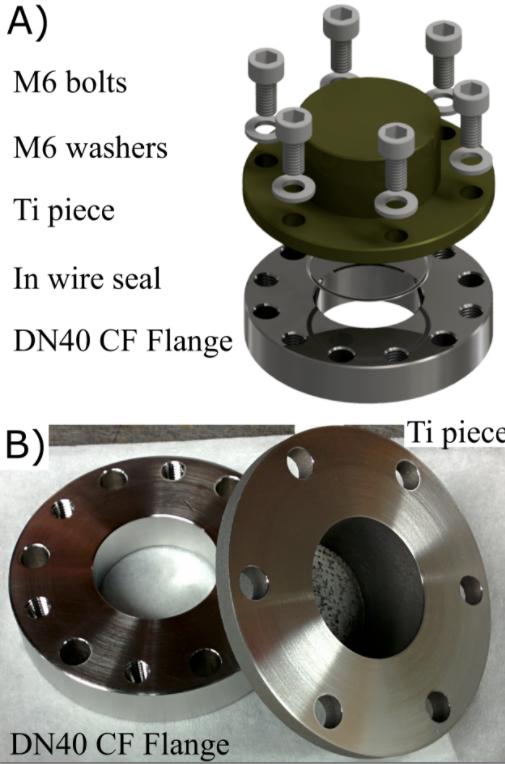}
\caption{A) Digital render of the vacuum flange structure with components for vacuum sealing. The test piece was designed with a top hat profile with a diameter of 70 mm and indium sealed onto a stainless steel flange. The stainless steel flange had a bore hole of 30 mm in the centre and 6 M6 tapped holes to apply pressure to the indium seal. Once sealed the DN40CF flange is attached to the chamber using a standard copper gasket. The holes in test piece are compatible with a DN40CF flange. B) The DN40 CF flange and test piece prior to Assembly.}
\label{fig:3DPrintFlange}
\end{figure}

The 3D-printed part was designed using CAD software and titanium (Ti) was chosen as it is a common vacuum material and readily 3D-printed. Furthermore, its non-magnetic properties are well-suited to cold atomic physics. The test piece was printed using direct metal laser sintering of alloy Ti-6Al-4V. The surface face of the test piece was smoothed with a milling machine to allow for indium sealing to the system and cleaned ultrasonically with acetone. The inside of the 3D-printed part was not smoothed. When connected to a vacuum the seal was done by indium wire with 1 mm diameter. The test piece and the method of sealing to a DN40CF can be seen in figure \ref{fig:3DPrintFlange}. 

\begin{figure}[h]
\centering
\includegraphics[scale=0.4]{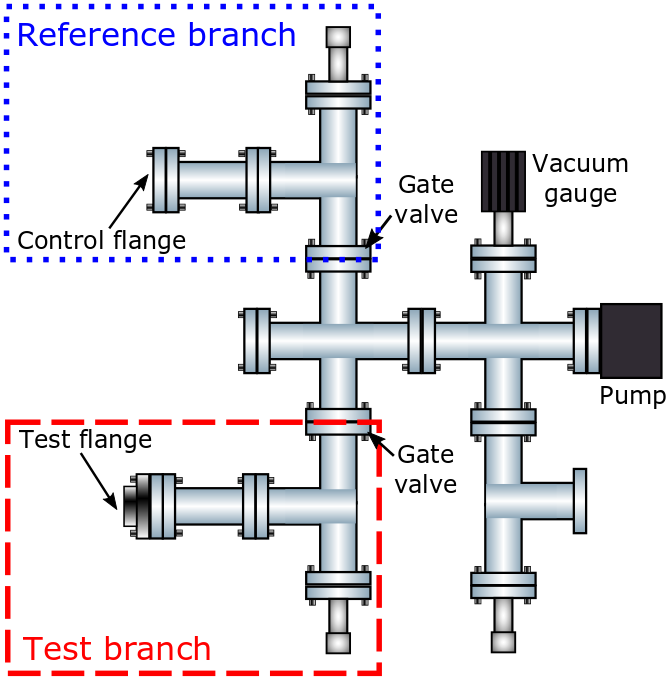}
\caption{The experimental set up for testing the performance of the 3D-printed flange compared to a commercial equivalent.}
\label{fig:3DFlangeExperimentalSetup}
\end{figure}

The experimental set up used to test the 3D-printed flange can be seen in figure \ref{fig:3DFlangeExperimentalSetup}. The set up consists of two branches which can be independently sealed off from each other and the rest of the chamber. The test branch contains the test piece, while the reference branch contains a control flange, produced by traditional methods (CFB70, MDC Vacuum LDC made of 304 stainless steel). The setup is pumped by an ion pump and the pumping speed for the two branches is the same due to the same conductance to the ion pump. 

\begin{figure}[h]
\centering
\includegraphics[scale=0.6]{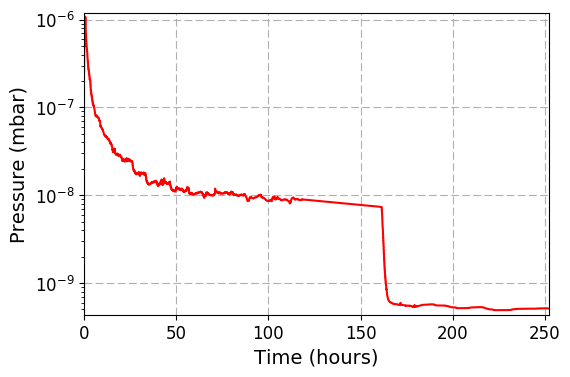}
\caption{Chamber pressure of the entire system vs time, after equalisation at a bake-out temperature of $ 130  \pm 1 ^\circ$C. The temperature for the bake out was kept below that of the melting point of indium ($\sim 157 ^\circ$C).  After 160 hours the bake out ended and the temperature of the system returned to room temperature ($\sim 21^\circ$C), and the pressure in the chamber reduced by roughly an order of magnitude.}
\label{fig:BakeOut}
\end{figure}

Pumping was provided by a triode configuration ion pump (Agilent technologies, Vaclon Plus 20 StarCell) and the pressure was measured using a cold cathode gauge (Pfeiffer, IKR 270). Moderate bake-out temperatures were achieved using heater tapes (SWH Series, Omega UK) and adjustable transformers (EA-STT 2000B-4.5A, Farnell), providing a stable temperature of $130 \pm 1$ $^\circ$C for 160 hours. The analogue output from the gauge controller was recorded every 0.5 s. The pressure during bakeout and cooling down can be seen in figure \ref{fig:BakeOut}. The data has been smoothed using a median filter to remove noise spikes. The ultimate pressure achieved after baking was 5$ \pm 0.5 \times 10^{-10}$ mbar. The same base pressure was achieved when both the reference and test branches were isolated from each other. From this it can be seen that the minimum achievable pressure in the vacuum system is not limited by the test flange

\begin{figure}[h]
\centering
\includegraphics[scale=0.6]{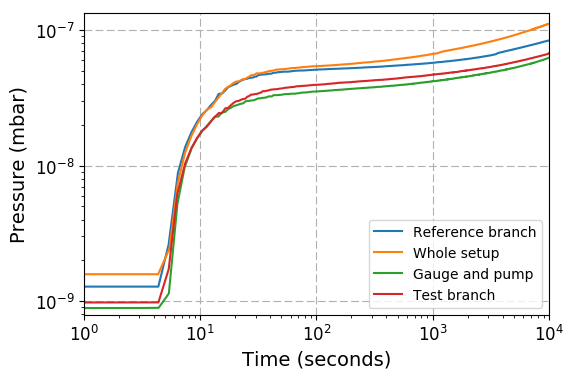}
\caption{Rate of rise curves for the different sections of the vacuum system. The pressure before the rate of rise curve is not representative of the minimum achievable pressures but instead of the pressure in the chamber before the test was carried out.}
\label{fig:PressureRiseTests}
\end{figure}

The performance of the test flange was assessed via pressure rise tests. The results of these tests can be seen in figure \ref{fig:PressureRiseTests}. The test branch was isolated from the reference branch and a pressure rise test preformed by turning the ion pump off (cyan trace in figure \ref{fig:PressureRiseTests}). Second, the reference branch of the setup was isolated and the same pressure rise test was performed for the test branch (red trace in figure \ref{fig:PressureRiseTests}). In addition, the pressure rise test was performed for the whole setup (blue trace in figure \ref{fig:PressureRiseTests}) and for the pump and gauge only (green trace in figure \ref{fig:PressureRiseTests}).

\begin{figure}[h]
\centering
\includegraphics[scale=0.6]{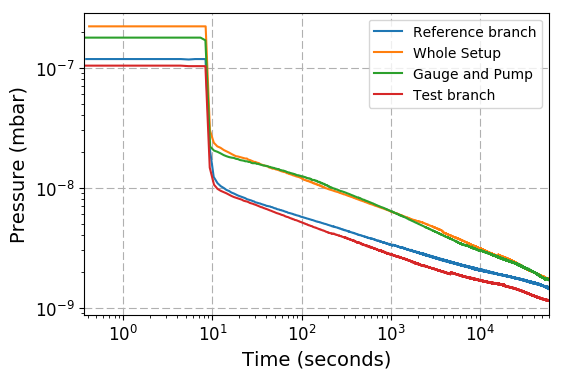}
\caption{The pressure for the different branches in the vacuum system as a function of time, when the chamber is pumping down.}
\label{fig:PressureDownTests}
\end{figure}

From the rate of rise  tests the out gassing rate for the reference branch was measured to be $5 \times 10^{-12}$ mbar l s$^{-1}$ cm$^{-2}$ while for the test branch it was measured to be $4 \times 10^{-12}$ mbar l s$^{-1}$ cm$^{-2}$, showing that the outgassing rate from the test flange does not exceed the control flange in the pressure ranges examined. The performance of the test flange and control flange during pump down is shown in figure \ref{fig:PressureDownTests} and measured by isolating each part of the system in turn and turning on the ion pump. The test flange and control flange both decreased in pressure by two orders of magnitude in 16.5 hours, reaching pressures of the order $1 \times 10^{-9}$ mbar. In addition to this the vacuum chamber has remained at UHV with no deterioration in performance over 2 years, showing great promise for the vacuum compatibility of 3D-printed titanium vacuum components and chambers. 

The pressure achieved in the system is not limited by the 3D-printed flange in the pressure range investigated, but rather the system out gassing due to a limited baking temperature of $130 \pm 1$ $^\circ$C, which is in turn limited by the use of an indium seal. If a 3D-printed knife edge could be realised, higher bake out pressures could be achieved and low pressure regimes could be investigated. It is possible that at lower pressures than studied here that the performance of the test and reference flange will deviate. However, the results demonstrated here show that 3D-printed titanium vacuum parts are suitable for use in the creation of UHV chambers and that no surface finishing is needed on the inside of the vacuum parts to achieve vacuum levels of interest for cold atom quantum sensors.

\section*{\label{sec:Con}Discussion}

The use of 3D-printing for the creation of both magnetic shielding and vacuum chambers has been demonstrated. Initial prototypes for 3D-printed magnetic shields show shielding factors within a factor of 3 of conventional approaches. While this may be enhanced further through optimisation of the fabrication and post-processing techniques, the main benefit that this brings is to enable the manufacture of more complex and compact shields, allowing a higher shielding factor per unit weight. This promises to drastically reduce the weight of quantum technology based sensors, where the magnetic shielding can occupy up to 50$\%$ of the weight.  The vacuum test pieces have shown that 3D-printed titanium structures are suitable for use as vacuum walls, as a sealing surface, and by association chambers in the pressure ranges examined. The test system reached base pressures of $5 \pm 0.5 \times 10^{-10}$ mbar and showed no signs of limiting the system. This has been achieved without post-processing or surface finishing of the internal walls. Combining the ability to 3D-print both magnetic shielding and vacuum systems will allow for a step change in the compactness and weight of quantum devices and facilitate greater systems integration. This potential for greater systems integration in cold atom systems is promising for applications where compactness is key, such as space based systems \cite{0264-9381-31-11-115010}. There is also the potential to benefit wider applications, in particular in sectors which require low weight shielding such as aerospace.

\section*{Methods}

\subsection*{\label{sec:Ack}Magnetic Shielding Production details}

The powder used to produce the magnetic shielding samples presented here was a Ni-5Mo-15Fe alloy (or permalloy-80), purchased from the TLS Technik company. All testing samples were produced on a  400-W powered Concept Laser M2 Cusing system, operating in an argon atmosphere using gas atomised powders in the size range 15-53 microns. Once printed special heat treatments are used to enhance the physical properties of  printed material. Two different types of heat treatment were applied to the samples produced by SLM manufacturing of permalloy-80.

\begin{figure}[h]
\centering
\includegraphics[scale=0.3]{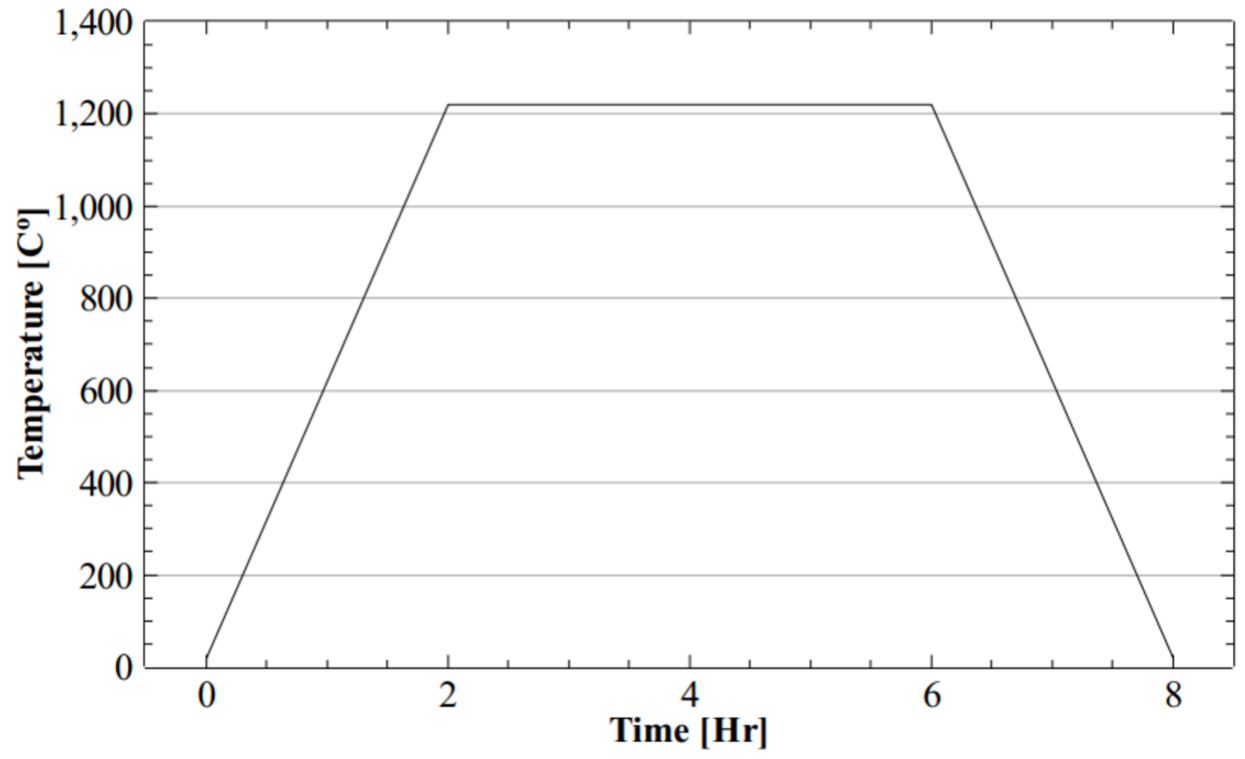}
\caption{Hot isostatic pressing process cycle in argon atmosphere at a pressure of 118 MPa, used for the 3D-printed sample in figure \ref{fig:OpenField} D.}
\label{fig:Hot isostatic pressing process}
\end{figure}

The process parameters for the first method, hot isostatic pressing (HIP), are shown in the time–temperature diagram
in figure \ref{fig:Hot isostatic pressing process}. The selected cycle for this design of experiments is typically used for 3D-printed materials of similar composition, predominantly to eliminate structural defects. 

\begin{figure}[h]
\centering
\includegraphics[scale=0.3]{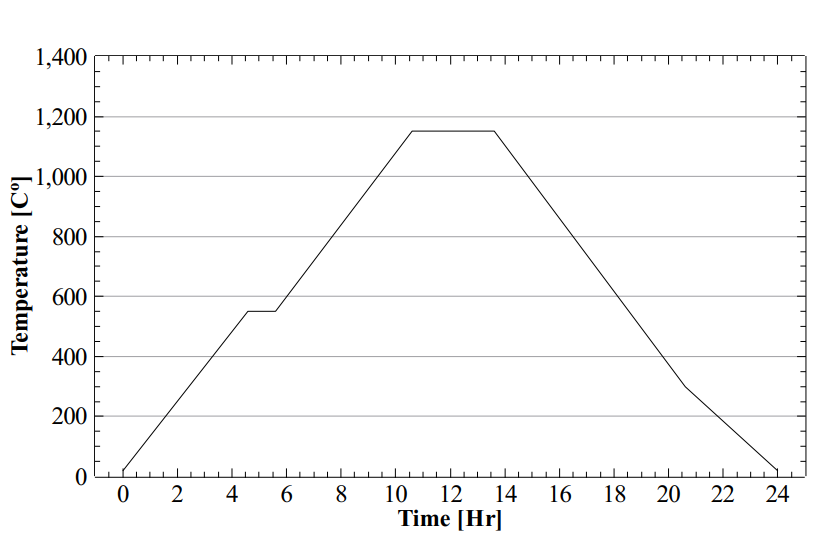}
\caption{Annealing treatment cycle in pure hydrogen atmosphere used for all the 3D-printed magnetic shield samples and mu metal shield.}
\label{fig:Annealing treatment cycle}
\end{figure}

The second method is annealing in pure dry hydrogen, which was undertaken at the Magnetic Shields Ltd. facilities. This cycle is used commercially by the company for bringing mu-metal shields to optimum magnetic performance as the final step after manufacturing. The relevant parameters are shown in figure \ref{fig:Annealing treatment cycle}.

\subsection*{Magnetic Shielding Measurements}

In preparation for testing the shielding performance, all the shields were demagnetised after production using a small degausser (Eclipse Magnetics, DA955 demagnetiser). The measurements of the magnetic shielding factors where performed using a Stefan Mayer 1-axis Fluxgate Magnetometer Fluxmaster which was fixed at the centre of the coils, while the shield was moved around it. This was achieved by a rail base that allowed the shield to slide from the one side to the other, through the coils, encompassing the sensor. The applied magnetic field strength in all experiments was 50 $\mu$T applied in the transverse orientation configuration.

\subsection*{3D-printed flange production details}

The 3D-printed test flange was prepared for printing using Solid Works and i.materialise.

\bibliography{Bib}

\begin{thebibliography}{10}
\expandafter\ifx\csname url\endcsname\relax
  \def\url#1{\texttt{#1}}\fi
\expandafter\ifx\csname urlprefix\endcsname\relax\def\urlprefix{URL }\fi
\expandafter\ifx\csname doiprefix\endcsname\relax\def\doiprefix{DOI }\fi
\providecommand{\bibinfo}[2]{#2}
\providecommand{\eprint}[2][]{\url{#2}}

\bibitem{BODDICE2017149}
\bibinfo{author}{Boddice, D.}, \bibinfo{author}{Metje, N.} \&
  \bibinfo{author}{Tuckwell, G.}
\newblock \bibinfo{journal}{\bibinfo{title}{Capability assessment and
  challenges for quantum technology gravity sensors for near surface
  terrestrial geophysical surveying}}.
\newblock {\emph{\JournalTitle{Journal of Applied Geophysics}}}
  \textbf{\bibinfo{volume}{146}}, \bibinfo{pages}{149 -- 159}
  (\bibinfo{year}{2017}).

\bibitem{dowling2003quantum}
\bibinfo{author}{Dowling, J.~P.} \& \bibinfo{author}{Milburn, G.~J.}
\newblock \bibinfo{journal}{\bibinfo{title}{Quantum technology: the second
  quantum revolution}}.
\newblock {\emph{\JournalTitle{Philosophical Transactions of the Royal Society
  of London A: Mathematical, Physical and Engineering Sciences}}}
  \textbf{\bibinfo{volume}{361}}, \bibinfo{pages}{1655--1674}
  (\bibinfo{year}{2003}).

\bibitem{keil2016fifteen}
\bibinfo{author}{Keil, M.} \emph{et~al.}
\newblock \bibinfo{journal}{\bibinfo{title}{Fifteen years of cold matter on the
  atom chip: promise, realizations, and prospects}}.
\newblock {\emph{\JournalTitle{Journal of modern optics}}}
  \textbf{\bibinfo{volume}{63}}, \bibinfo{pages}{1840--1885}
  (\bibinfo{year}{2016}).

\bibitem{Kovachy2015}
\bibinfo{author}{Kovachy, T.} \emph{et~al.}
\newblock \bibinfo{journal}{\bibinfo{title}{Quantum superposition at the
  half-metre scale}}.
\newblock {\emph{\JournalTitle{Nature}}} \textbf{\bibinfo{volume}{528}},
  \bibinfo{pages}{530--533} (\bibinfo{year}{2015}).

\bibitem{BloomClock}
\bibinfo{author}{Bloom, B.~J.} \emph{et~al.}
\newblock \bibinfo{journal}{\bibinfo{title}{An optical lattice clock with
  accuracy and stability at the 10-18 level}}.
\newblock {\emph{\JournalTitle{Nature}}} \textbf{\bibinfo{volume}{506}},
  \bibinfo{pages}{71--75}.

\bibitem{Sander:12}
\bibinfo{author}{Sander, T.~H.} \emph{et~al.}
\newblock \bibinfo{journal}{\bibinfo{title}{Magnetoencephalography with a
  chip-scale atomic magnetometer}}.
\newblock {\emph{\JournalTitle{Biomed. Opt. Express}}}
  \textbf{\bibinfo{volume}{3}}, \bibinfo{pages}{981--990}
  (\bibinfo{year}{2012}).

\bibitem{hinton2017portable}
\bibinfo{author}{Hinton, A.} \emph{et~al.}
\newblock \bibinfo{journal}{\bibinfo{title}{A portable magneto-optical trap
  with prospects for atom interferometry in civil engineering}}.
\newblock {\emph{\JournalTitle{Phil. Trans. R. Soc. A}}}
  \textbf{\bibinfo{volume}{375}}, \bibinfo{pages}{20160238}
  (\bibinfo{year}{2017}).

\bibitem{wu2014investigation}
\bibinfo{author}{Wu, B.} \emph{et~al.}
\newblock \bibinfo{journal}{\bibinfo{title}{The investigation of a
  $\mu$gal-level cold atom gravimeter for field applications}}.
\newblock {\emph{\JournalTitle{Metrologia}}} \textbf{\bibinfo{volume}{51}},
  \bibinfo{pages}{452} (\bibinfo{year}{2014}).

\bibitem{hauth2013first}
\bibinfo{author}{Hauth, M.} \emph{et~al.}
\newblock \bibinfo{journal}{\bibinfo{title}{First gravity measurements using
  the mobile atom interferometer gain}}.
\newblock {\emph{\JournalTitle{Applied Physics B}}}
  \textbf{\bibinfo{volume}{113}}, \bibinfo{pages}{49--55}
  (\bibinfo{year}{2013}).

\bibitem{menoret2011dual}
\bibinfo{author}{M{\'e}noret, V.} \emph{et~al.}
\newblock \bibinfo{journal}{\bibinfo{title}{Dual-wavelength laser source for
  onboard atom interferometry}}.
\newblock {\emph{\JournalTitle{Optics Letters}}} \textbf{\bibinfo{volume}{36}},
  \bibinfo{pages}{4128--4130} (\bibinfo{year}{2011}).

\bibitem{politi2008silica}
\bibinfo{author}{Politi, A.}, \bibinfo{author}{Cryan, M.~J.},
  \bibinfo{author}{Rarity, J.~G.}, \bibinfo{author}{Yu, S.} \&
  \bibinfo{author}{O'brien, J.~L.}
\newblock \bibinfo{journal}{\bibinfo{title}{Silica-on-silicon waveguide quantum
  circuits}}.
\newblock {\emph{\JournalTitle{Science}}} \textbf{\bibinfo{volume}{320}},
  \bibinfo{pages}{646--649} (\bibinfo{year}{2008}).

\bibitem{sewell2010atom}
\bibinfo{author}{Sewell, R.} \emph{et~al.}
\newblock \bibinfo{journal}{\bibinfo{title}{Atom chip for {BEC}
  interferometry}}.
\newblock {\emph{\JournalTitle{Journal of Physics B: Atomic, Molecular and
  Optical Physics}}} \textbf{\bibinfo{volume}{43}}, \bibinfo{pages}{051003}
  (\bibinfo{year}{2010}).

\bibitem{PhysRevLett.110.093602}
\bibinfo{author}{M\"untinga, H.} \emph{et~al.}
\newblock \bibinfo{journal}{\bibinfo{title}{Interferometry with bose-einstein
  condensates in microgravity}}.
\newblock {\emph{\JournalTitle{Phys. Rev. Lett.}}}
  \textbf{\bibinfo{volume}{110}}, \bibinfo{pages}{093602}
  (\bibinfo{year}{2013}).

\bibitem{PhysRevA.96.033636}
\bibinfo{author}{Imhof, E.} \emph{et~al.}
\newblock \bibinfo{journal}{\bibinfo{title}{Two-dimensional grating
  magneto-optical trap}}.
\newblock {\emph{\JournalTitle{Phys. Rev. A}}} \textbf{\bibinfo{volume}{96}},
  \bibinfo{pages}{033636} (\bibinfo{year}{2017}).

\bibitem{Wonderfulwidgets}
\bibinfo{journal}{\bibinfo{title}{Wonderful widgets}}.
\newblock {\emph{\JournalTitle{{The Economist}}}}  (\bibinfo{year}{2015}).
\newblock
  \urlprefix\url{https://www.economist.com/news/technology-quarterly/21662653-components-become-more-elegant-software-produces-most-efficient}.

\bibitem{TheEconomist}
\bibinfo{journal}{\bibinfo{title}{3{D} printing and clever computers could
  revolutionise construction}}.
\newblock {\emph{\JournalTitle{{The Economist}}}}  (\bibinfo{year}{2017}).
\newblock
  \urlprefix\url{https://www.economist.com/news/science-and-technology/21722820-think-spiderweb-floors-denser-skyscrapers-and-ultra-thin-bridges-3d-printing-and}.

\bibitem{doi:10.1063/1.4937614}
\bibinfo{author}{Zhang, G.~H.}, \bibinfo{author}{Braverman, B.},
  \bibinfo{author}{Kawasaki, A.} \& \bibinfo{author}{Vuletić, V.}
\newblock \bibinfo{journal}{\bibinfo{title}{Note: Fast compact laser shutter
  using a direct current motor and three-dimensional printing}}.
\newblock {\emph{\JournalTitle{Review of Scientific Instruments}}}
  \textbf{\bibinfo{volume}{86}}, \bibinfo{pages}{126105}
  (\bibinfo{year}{2015}).

\bibitem{Ling}
\bibinfo{author}{Li, L.} \emph{et~al.}
\newblock \bibinfo{journal}{\bibinfo{title}{Big area additive manufacturing of
  high performance bonded {N}d{F}e{B} magnets}}.
\newblock {\emph{\JournalTitle{Scientific Reports}}}
  \textbf{\bibinfo{volume}{6}} (\bibinfo{year}{2016}).

\bibitem{doi:10.1063/1.4964856}
\bibinfo{author}{Huber, C.} \emph{et~al.}
\newblock \bibinfo{journal}{\bibinfo{title}{3d print of polymer bonded
  rare-earth magnets, and 3d magnetic field scanning with an end-user 3d
  printer}}.
\newblock {\emph{\JournalTitle{Applied Physics Letters}}}
  \textbf{\bibinfo{volume}{109}}, \bibinfo{pages}{162401}
  (\bibinfo{year}{2016}).

\bibitem{zhou2017design}
\bibinfo{author}{Zhou, Y.} \emph{et~al.}
\newblock \bibinfo{journal}{\bibinfo{title}{Design of magneto-optical traps for
  additive manufacture by 3{D} printing}}.
\newblock {\emph{\JournalTitle{arXiv preprint arXiv:1704.00430}}}
  (\bibinfo{year}{2017}).

\bibitem{saint20173d}
\bibinfo{author}{Saint, R.} \emph{et~al.}
\newblock \bibinfo{journal}{\bibinfo{title}{3{D}-printed components for quantum
  devices}}.
\newblock {\emph{\JournalTitle{arXiv preprint arXiv:1704.01813}}}
  (\bibinfo{year}{2017}).

\bibitem{tong2016advanced}
\bibinfo{author}{Tong, X.~C.}
\newblock \emph{\bibinfo{title}{Advanced materials and design for
  electromagnetic interference shielding}} (\bibinfo{publisher}{CRC press},
  \bibinfo{year}{2016}).

\bibitem{doi:10.1063/1.4720943}
\bibinfo{author}{Dickerson, S.} \emph{et~al.}
\newblock \bibinfo{journal}{\bibinfo{title}{A high-performance magnetic shield
  with large length-to-diameter ratio}}.
\newblock {\emph{\JournalTitle{Review of Scientific Instruments}}}
  \textbf{\bibinfo{volume}{83}}, \bibinfo{pages}{065108}
  (\bibinfo{year}{2012}).

\bibitem{PhysRevLett.90.160401}
\bibinfo{author}{Hornberger, K.} \emph{et~al.}
\newblock \bibinfo{journal}{\bibinfo{title}{Collisional decoherence observed in
  matter wave interferometry}}.
\newblock {\emph{\JournalTitle{Phys. Rev. Lett.}}}
  \textbf{\bibinfo{volume}{90}}, \bibinfo{pages}{160401}
  (\bibinfo{year}{2003}).

\bibitem{das1998producing}
\bibinfo{author}{Das, S.}, \bibinfo{author}{Wohlert, M.},
  \bibinfo{author}{Beaman, J.~J.} \& \bibinfo{author}{Bourell, D.~L.}
\newblock \bibinfo{journal}{\bibinfo{title}{Producing metal parts with
  selective laser sintering/hot isostatic pressing}}.
\newblock {\emph{\JournalTitle{JOM Journal of The Minerals, Metals and
  Materials Society}}} \textbf{\bibinfo{volume}{50}}, \bibinfo{pages}{17--20}
  (\bibinfo{year}{1998}).

\bibitem{povilus2014vacuum}
\bibinfo{author}{Povilus, A.~P.}, \bibinfo{author}{Wurden, C.~J.},
  \bibinfo{author}{Vendeiro, Z.}, \bibinfo{author}{Baquero-Ruiz, M.} \&
  \bibinfo{author}{Fajans, J.}
\newblock \bibinfo{journal}{\bibinfo{title}{Vacuum compatibility of
  3{D}-printed materials}}.
\newblock {\emph{\JournalTitle{Journal of Vacuum Science \& Technology A:
  Vacuum, Surfaces, and Films}}} \textbf{\bibinfo{volume}{32}},
  \bibinfo{pages}{033001} (\bibinfo{year}{2014}).

\bibitem{parimi2014additive}
\bibinfo{author}{Parimi, L.~L.}
\newblock \emph{\bibinfo{title}{Additive manufacturing of nickel based
  superalloys for aerospace applications}}.
\newblock Ph.D. thesis, \bibinfo{school}{University of Birmingham}
  (\bibinfo{year}{2014}).

\bibitem{shishkovsky2015comparison}
\bibinfo{author}{Shishkovsky, I.~V.}, \bibinfo{author}{Nazarov, A.~P.},
  \bibinfo{author}{Kotoban, D.~V.} \& \bibinfo{author}{Kakovkina, N.~G.}
\newblock \bibinfo{title}{Comparison of additive technologies for gradient
  aerospace part fabrication from nickel-based superalloys}.
\newblock In \emph{\bibinfo{booktitle}{Superalloys}}
  (\bibinfo{publisher}{InTech}, \bibinfo{year}{2015}).

\bibitem{mikler2017laser}
\bibinfo{author}{Mikler, C.} \emph{et~al.}
\newblock \bibinfo{journal}{\bibinfo{title}{Laser additive processing of
  {N}i-{F}e-{V} and {N}i-{F}e-{M}o permalloys: Microstructure and magnetic
  properties}}.
\newblock {\emph{\JournalTitle{Materials Letters}}}
  \textbf{\bibinfo{volume}{192}}, \bibinfo{pages}{9--11}
  (\bibinfo{year}{2017}).

\bibitem{shishkovsky2016peculiarities}
\bibinfo{author}{Shishkovsky, I.} \& \bibinfo{author}{Saphronov, V.}
\newblock \bibinfo{journal}{\bibinfo{title}{Peculiarities of selective laser
  melting process for permalloy powder}}.
\newblock {\emph{\JournalTitle{Materials Letters}}}
  \textbf{\bibinfo{volume}{171}}, \bibinfo{pages}{208--211}
  (\bibinfo{year}{2016}).

\bibitem{carter2014influence}
\bibinfo{author}{Carter, L.~N.}, \bibinfo{author}{Martin, C.},
  \bibinfo{author}{Withers, P.~J.} \& \bibinfo{author}{Attallah, M.~M.}
\newblock \bibinfo{journal}{\bibinfo{title}{The influence of the laser scan
  strategy on grain structure and cracking behaviour in slm powder-bed
  fabricated nickel superalloy}}.
\newblock {\emph{\JournalTitle{Journal of Alloys and Compounds}}}
  \textbf{\bibinfo{volume}{615}}, \bibinfo{pages}{338--347}
  (\bibinfo{year}{2014}).

\bibitem{paperno2000new}
\bibinfo{author}{Paperno, E.}, \bibinfo{author}{Koide, H.} \&
  \bibinfo{author}{Sasada, I.}
\newblock \bibinfo{journal}{\bibinfo{title}{A new estimation of the axial
  shielding factors for multishell cylindrical shields}}.
\newblock {\emph{\JournalTitle{Journal of Applied Physics}}}
  \textbf{\bibinfo{volume}{87}}, \bibinfo{pages}{5959--5961}
  (\bibinfo{year}{2000}).

\bibitem{Carter}
\bibinfo{author}{Carter, L.~N.} \emph{et~al.}
\newblock \bibinfo{journal}{\bibinfo{title}{Process optimisation of selective
  laser melting using energy density model for nickel based superalloys}}.
\newblock {\emph{\JournalTitle{Materials Science and Technology}}}
  \textbf{\bibinfo{volume}{32}}, \bibinfo{pages}{657--661}
  (\bibinfo{year}{2016}).

\bibitem{o1999modern}
\bibinfo{author}{O'Handley, R.}
\newblock \emph{\bibinfo{title}{Modern Magnetic Materials: Principles and
  Applications}} (\bibinfo{publisher}{Wiley}, \bibinfo{year}{1999}).

\bibitem{lampman1997weld}
\bibinfo{author}{Lampman, S.} \& \bibinfo{author}{International, A.}
\newblock \emph{\bibinfo{title}{Weld Integrity and Performance: A Source Book
  Adapted from ASM International Handbooks, Conference Proceedings, and
  Technical Books}}.
\newblock ASM Handbook Series (\bibinfo{publisher}{ASM International},
  \bibinfo{year}{1997}).

\bibitem{g}
\bibinfo{author}{Larker, H.~T.} \& \bibinfo{author}{Larker, R.}
\newblock \emph{\bibinfo{title}{Hot Isostatic Pressing}}
  (\bibinfo{publisher}{Wiley-VCH Verlag GmbH \& Co. KGaA},
  \bibinfo{year}{2006}).

\bibitem{gans2014vacuum}
\bibinfo{author}{Gans, A.~R.}, \bibinfo{author}{Jobbins, M.~M.},
  \bibinfo{author}{Lee, D.~Y.} \& \bibinfo{author}{Alex~Kandel, S.}
\newblock \bibinfo{journal}{\bibinfo{title}{Vacuum compatibility of silver and
  titanium parts made using three-dimensional printing}}.
\newblock {\emph{\JournalTitle{Journal of Vacuum Science \& Technology A:
  Vacuum, Surfaces, and Films}}} \textbf{\bibinfo{volume}{32}},
  \bibinfo{pages}{023201} (\bibinfo{year}{2014}).

\bibitem{1742-6596-874-1-012097}
\bibinfo{author}{Jenzer, S.} \emph{et~al.}
\newblock \bibinfo{journal}{\bibinfo{title}{Study of the suitability of 3{D}
  printing for ultra-high vacuum applications}}.
\newblock {\emph{\JournalTitle{Journal of Physics: Conference Series}}}
  \textbf{\bibinfo{volume}{874}}, \bibinfo{pages}{012097}
  (\bibinfo{year}{2017}).

\bibitem{0264-9381-31-11-115010}
\bibinfo{author}{Aguilera, D.~N.} \emph{et~al.}
\newblock \bibinfo{journal}{\bibinfo{title}{{STE-QUEST}—test of the
  universality of free fall using cold atom interferometry}}.
\newblock {\emph{\JournalTitle{Classical and Quantum Gravity}}}
  \textbf{\bibinfo{volume}{31}}, \bibinfo{pages}{115010}
  (\bibinfo{year}{2014}).

\end{thebibliography}

\section*{Acknowledgements}

The authors would like to acknowledge funding from EPSRC through grant EP/M013294, DSTL through contract DSTLX-1000095040, and the financial support of the Future and Emerging Technologies (FET) programme within the 7th Framework Programme for 
Research of the European Commission, under FET grant number: FP7-ICT-601180.

\section*{Author contributions statement}
M.H, K.B and V.B conceived the experiments and supervised the research. J.V, G.V and P.P tested and characterised the vacuum and magnetic shielding pieces and analysed the results. J.Z and Y.G fabricated the 3D-printed shield and characterised their structure, supervised by M.M.A. L.B and D.W performed the heat treatments and made the control shield. J.V wrote the paper with help from all the other co-authors. All authors reviewed the manuscript. 

\section*{Competing financial interests}

The authors declare that they have no competing interests.

\section*{Data availability statement}

Supporting data can be provided if requested.

\end{document}